\documentclass[letterpaper, 12pt]{article}[2000/05/19]
\usepackage[english]{babel}
\usepackage{amsfonts,amsmath,amssymb,amsthm,latexsym,amscd,mathrsfs}
\usepackage{ifthen,cite}
\usepackage[bookmarksnumbered=true]{hyperref}

\hypersetup{pdfpagetransition={Split}}

\newcommand{\evenhead}{AuthorNameForHeading \ name}
\newcommand{\oddhead}{ShortArticleName \ name}
\newcommand{\theArticleName}{Article name}

\newcommand{\FirstPageHeading}[1]{\thispagestyle{empty}%
\noindent\raisebox{0pt}[0pt][0pt]{\makebox[\textwidth]{\protect\footnotesize \sf }}\par}

\newcommand{\ArticleName}[1]{\renewcommand{\theArticleName}{#1}\vspace{-2mm}\par\noindent {\LARGE\bf  #1\par}}
\newcommand{\Author}[1]{\vspace{5mm}\par\noindent {\Large  #1\par} \par\vspace{2mm}\par}
\newcommand{\Address}[1]{\vspace{2mm}\par\noindent {\it #1} \par}
\newcommand{\Abstract}[1]{\vspace{6mm}\par\noindent\hspace*{10mm}
\parbox{140mm}{\small {\bf Abstract.} #1}\par}
\newcommand{\Keywords}[1]{\vspace{3mm}\par\noindent\hspace*{10mm}
\parbox{140mm}{\small {\bf Key words:} \rm #1}\par}
\newcommand{\Classification}[1]{\vspace{3mm}\par\noindent\hspace*{10mm}
\parbox{140mm}{\small {\it 2000 Mathematics Subject Classification:} \rm #1}\vspace{3mm}\par}
\newcommand{\ShortArticleName}[1]{\renewcommand{\oddhead}{#1}}
\newcommand{\AuthorNameForHeading}[1]{\renewcommand{\evenhead}{#1}}

\long\def\@makecaption#1#2{
  \sbox\@tempboxa{\small \textbf{#1.}\ \ #2}%
  \ifdim \wd\@tempboxa >\hsize
    {\small \textbf{#1.}\ \ #2}\par \else
    \global \@minipagefalse
    \hb@xt@\hsize{\hfil\box\@tempboxa\hfil}%
  \fi \vskip\belowcaptionskip}

\def\numberwithin#1#2{\@ifundefined{c@#1}{\@nocounterr{#1}}{%
  \@ifundefined{c@#2}{\@nocnterr{#2}}{%
  \@addtoreset{#1}{#2}%
  \toks@\@xp\@xp\@xp{\csname the#1\endcsname}%
  \@xp\xdef\csname the#1\endcsname
    {\@xp\@nx\csname the#2\endcsname.\the\toks@}}}}

{\theoremstyle{definition}

}

\begin{document}

\FirstPageHeading{V.I. Gerasimenko and Zh.A. Tsvir}

\ShortArticleName{Quantum kinetic equations}

\AuthorNameForHeading{V.I. Gerasimenko, Zh.A. Tsvir}

\ArticleName{Quantum Kinetic Equations\\ of Many-Particle Systems in Condensed States}

\Author{V.I. Gerasimenko$^\ast$\footnote{E-mail: \emph{gerasym@imath.kiev.ua}}
        and Zh.A. Tsvir$^\ast$$^\ast$\footnote{E-mail: \emph{Zhanna.Tsvir@simcorp.com}}}

\Address{$^\ast$\hspace*{1mm}Institute of Mathematics of NAS of Ukraine,\\
         \hspace*{3mm}3, Tereshchenkivs'ka Str.,\\
         \hspace*{3mm}01601 Kyiv-4, Ukraine}

\Address{$^\ast$$^\ast$Taras Shevchenko National University of Kyiv,\\
    \hspace*{4mm}Department of Mechanics and Mathematics,\\
    \hspace*{4mm}2, Academician Glushkov Av.,\\
    \hspace*{4mm}03187 Kyiv, Ukraine}

\bigskip

\Abstract{This paper is devoted to the description of the evolution of states of quantum
many-particle systems within the framework of a one-particle density operator, which enables
to construct the kinetic equations in scaling limits in the presence of correlations of particle
states at initial time, for instance, correlations characterizing the condensed states.
}

\bigskip

\vspace{2pc}
\Classification{35Q40; 47J35; 47H20; 82C10; 82C22.}

\makeatletter
\renewcommand{\@evenhead}{
\hspace*{-3pt}\raisebox{-7pt}[\headheight][0pt]{\vbox{\hbox to \textwidth {\thepage \hfil \evenhead}\vskip4pt \hrule}}}
\renewcommand{\@oddhead}{
\hspace*{-3pt}\raisebox{-7pt}[\headheight][0pt]{\vbox{\hbox to \textwidth {\oddhead \hfil \thepage}\vskip4pt \hrule}}}
\renewcommand{\@evenfoot}{}
\renewcommand{\@oddfoot}{}
\makeatother

\newpage
\vphantom{math}
\protect\tableofcontents

\vspace{0.5cm}

\section{Introduction}
Recently the considerable advance in the rigorous derivation of quantum kinetic equations
in the mean field scaling limit \cite{Sp80}, such as the nonlinear Schr\"{o}dinger equation
and the Gross-Pitaevskii equation \cite{AA,AGT,BGGM1,ESchY2,EShY10,ESch,GMM,M1,S-R,Sp07} is
observed. In the scaling limit approach to the construction of kinetic equations one of the
essential assumptions is that initial data has to satisfy a chaos property, i.e. the initial
state is completely determined by a sequence of products of the one-particle density operators
\cite{Sp80},\cite{CGP97}. At the same time it is well known that, for instance, the equilibrium
state of the Bose condensate is characterized by correlations of particles in contrast to the
gaseous state \cite{BQ}. Thus, the validation of quantum kinetic equations in case of presence
of correlations at initial time, in particular for condensed states, remains an actual problem so far.

In the paper we consider the problem of potentialities inherent in the description of the evolution
of states of many-particle systems in terms of a one-particle density operator in case of presence
of correlations of particles at initial time. We prove that, if initial data is completely
specified by a one-particle marginal density operator and initial correlations, then all possible
states of infinite-particle systems at arbitrary moment of time can be described without any
approximations within the framework of a one-particle density operator governed by the kinetic
equation, which coefficients are determined by the initial correlations. Then the actual kinetic
equations that describe the evolution of interacting particles in the condensed states, are
derived in appropriate scaling limits on the basis of the established quantum kinetic equation.

We consider a quantum system of a non-fixed (nonequilibrium grand canonical ensemble) number of
identical (spinless) particles obeying Maxwell-Boltzmann statistics in the space $\mathbb{R}^{3}$.
We will use units where $h={2\pi\hbar}=1$ is a Planck constant, and $m=1$ is the mass of particles.

Let $\mathcal{H}$ be a one-particle Hilbert space, then the $n$-particle space $\mathcal{H}_n$,
is a tensor product of $n$ Hilbert spaces $\mathcal{H}$. The Hamiltonian $H_{n}$ of $n$-particle
system is a self-adjoint operator with domain $\mathcal{D}(H_{n})\subset\mathcal{H}_{n}$:
\begin{eqnarray}\label{H}
    &&H_{n}=\sum\limits_{i=1}^{n}K(i)+\epsilon\sum\limits_{i_{1}<i_{2}=1}^{n}\Phi(i_{1},i_{2}),
\end{eqnarray}
where $K(i)$ is the operator of a kinetic energy of the $ith$ particle, $\Phi(i_{1},i_{2})$ is the
operator of a two-body interaction potential and $\epsilon>0$ is a scaling parameter. The operator
$K(i)$ acts on functions $\psi_n$, that belong to the subspace
$L^{2}_{0}(\mathbb{R}^{3n})\subset\mathcal{D}(H_n)\subset L^{2}(\mathbb{R}^{3n})$ of infinitely
differentiable functions with compact supports according to the formula:
$K(i)\psi_n=-\frac{1}{2}\Delta_{q_i}\psi_n$. Correspondingly we have:
$\Phi(i_{1},i_{2})\psi_{n}=\Phi(q_{i_{1}},q_{i_{2}})\psi_{n}$, and we assume that the function
$\Phi(q_{i_{1}},q_{i_{2}})$ is translation-invariant bounded function which is symmetric with respect
to permutations of arguments.

Let $\mathfrak{L}^{1}(\mathcal{H}_{n})$ be the space of trace class operators
$f_{n}\equiv f_{n}(1,\ldots,n)\in\mathfrak{L}^{1}(\mathcal{H}_{n})$ that satisfy the symmetry
condition: $f_{n}(1,\ldots,n)=f_{n}(i_{1},\ldots,i_{n})$ for arbitrary $(i_{1},\ldots,i_{n})\in(1,\ldots,n)$,
and equipped with the norm: $\|f_{n}\|_{\mathfrak{L}^{1}(\mathcal{H}_{n})}=
\mathrm{Tr}_{1,\ldots,n}|f_{n}(1,\ldots,n)|$, where $\mathrm{Tr}_{1,\ldots,n}$ are partial traces over
$1,\ldots,n$ particles. We denote by $\mathfrak{L}^{1}_0(\mathcal{H}_{n})$ the everywhere dense set of
finite sequences of degenerate operators with infinitely differentiable kernels with compact supports.
Let $\mathfrak{L}(\mathcal{H}_n)$ be the space of bounded operators defined on the Hilbert space $\mathcal{H}_n$.

On the space $\mathfrak{L}^{1}(\mathcal{F}_\mathcal{H})$ the one-parameter family of operators
\begin{eqnarray}\label{groupG}
    &&\mathcal{G}_{n}(-t)f_n\doteq e^{-itH_{n}}f_n\,e^{itH_{n}}.
\end{eqnarray}
is defined and it is an isometric strongly continuous group which preserves positivity and
self-adjointness of operators. For $f\in\mathfrak{L}^{1}_{0}(\mathcal{H}_{n})$ there exists
the infinitesimal generator of this group
\begin{eqnarray}\label{infOper}
    &&\lim\limits_{t\rightarrow 0}\frac{1}{t}\big(\mathcal{G}_{n}(-t)f_{n}-f_{n}\big)
       =-i(H_{n}f_{n}-f_{n}H_{n})\doteq-\mathcal{N}_{n}f_{n},
\end{eqnarray}
where $H_{n}$ is Hamiltonian (\ref{H}) and the operator: $-i(H_{n}f_{n}-f_{n}H_{n})$ is defined
on the domain $\mathcal{D}(H_{n})\subset\mathcal{H}_{n}$.

Let us denote by $\{Y\}$ the set consisting of one element $Y\equiv(1,\ldots,s)$. We define the
$(1+n)th$-order ($n\geq0$) cumulant of groups of operators (\ref{groupG}) as follows
\begin{eqnarray}\label{cumulant}
   &&\hskip-5mm\mathfrak{A}_{1+n}(-t,\{Y\},X\setminus Y)=
       \sum\limits_{\mathrm{P}:(\{Y\},\,X\setminus Y)
       ={\bigcup\limits}_i X_i}(-1)^{|\mathrm{P}|-1}(|\mathrm{P}|-1)!
     \prod_{X_i\subset\mathrm{P}}\mathcal{G}_{|\theta(X_i)|}(-t,\theta(X_i)),
\end{eqnarray}
where $\{Y\}$ is the set consisting of one element $Y=(1,\ldots,s)$,
${\sum\limits}_\mathrm{P}$ is the sum over all possible partitions $\mathrm{P}$ of the set
$(\{Y\},X\setminus Y)=(\{Y\},s+1,\ldots,s+n)$ into $|\mathrm{P}|$ nonempty mutually disjoint subsets
$X_i\subset(\{Y\},X\setminus Y)$ and $\theta: (\{Y\},X\setminus Y)\rightarrow X$
is the declasterization mapping.

We will consider initial state which is given by the following sequence of marginal density operators
\begin{eqnarray*}
   &&\hskip-5mmF(t)|_{t=0}=\big(1,F_1^0(1),g_{2}(1,2)F_1^0(1)F_1^0(2),\ldots,g_{n}(1,...,n)
       \prod_{i=1}^{n}F_1^0(i),\ldots\big),
\end{eqnarray*}
where the bounded operators $g_{n}\in\mathfrak{L}(\mathcal{H}_n),\,n\geq2$, are specified
initial correlations. Such initial data is typical for the condensed states of quantum gases,
for example, the equilibrium state of the Bose condensate satisfies the weakening of correlation
condition with the correlations which characterize the condensed state \cite{BQ}. We remark that
nonequilibrium dynamics of correlations was constructed in \cite{GP}.

The evolution of all possible states is described by the sequence of marginal density operators
$F(t)=\big(1,F_1(t,1),\ldots,F_s(t,1,\ldots,s),\ldots\big)$ governed by the initial-value problem
of the quantum BBGKY hierarchy
\begin{eqnarray}
 \label{BBGKY}
   &&\hskip-5mm\frac{d}{dt}F_{s}(t,Y)=\big(\sum\limits_{j=1}^{s}(-\mathcal{N}(j))+
       \epsilon\sum\limits_{j_1<j_2=1}^{s}(-\mathcal{N}_{\mathrm{int}}(j_1,j_2))\big)F_{s}(t,Y)+\nonumber \\
   &&\hskip-5mm+\epsilon\,\sum\limits_{j=1}^{s}
       \mathrm{Tr}_{s+1}(-\mathcal{N}_{\mathrm{int}}(j,s+1))F_{s+1}(t,Y,s+1),\\
       \nonumber \\
 \label{BBGKYi}
   &&\hskip-5mmF_{s}(t,Y)|_{t=0}=g_{s}(Y)\prod_{i=1}^{s}F_1^0(i),\quad s\geq 1,
\end{eqnarray}
where on $f_n\in\mathfrak{L}_{0}^1(\mathcal{H}_n)\subset\mathfrak{L}^1(\mathcal{H}_n)$ the operators
$(-\mathcal{N}(j))$ and $(-\mathcal{N}_{\mathrm{int}}(j_1,j_{2}))$ are defined by the corresponding
formula
\begin{eqnarray}\label{comst}
   &&(-\mathcal{N}(j))f_n\doteq -i(K(j)f_n-f_n K(j)),\\
   &&(-\mathcal{N}_{\mathrm{int}})(j_1,j_{2})f_n\doteq -
       i(\Phi(j_1,j_{2})f_n-f_n\Phi(j_1,j_{2})).\nonumber
\end{eqnarray}

If $\|F_{1}^0\|_{\mathfrak{L}^{1}(\mathcal{H})}<e^{-1}$ and
${\max}_{n\geq2}\|g_{n}\|_{\mathfrak{L}(\mathcal{H}_n)}<\infty$, then for $t\in\mathbb{R}$
a nonperturbative solution of the Cauchy problem (\ref{BBGKY})-(\ref{BBGKYi}) is given by
the expansion \cite{GT}
\begin{eqnarray}\label{RozvBBGKY}
   &&\hskip-5mmF_{s}(t,Y)=\sum\limits_{n=0}^{\infty}\frac{1}{n!}\,\mathrm{Tr}_{s+1,\ldots,{s+n}}\,
       \mathfrak{A}_{1+n}(-t,\{Y\},\,X\setminus Y)g_{1+n}(\{Y\},\,X\setminus Y)\prod_{i=1}^{s+n}F_1^0(i),
\end{eqnarray}
where $\mathfrak{A}_{1+n}(-t)$ is the $(1+n)th$-order ($n\geq0$) cumulant (\ref{cumulant}) of groups
of operators (\ref{groupG}) and $g_{1+n}(\{Y\},\,X\setminus Y)$ is the correlation operator of
clusters of particles \cite{GP}, in particular case $n=0$ it has the form: $g_{1+0}(t,\{Y\})=
{\sum}_{\mathrm{P}:\,Y=\bigcup_{i} X_{i}}{\prod}_{X_{i}\subset\mathrm{P}}g_{|X_{i}|}(t,X_{i})$.

In the paper \cite{GT} it was proved that, if initial data is completely determined by a one-particle
marginal density operator, i.e. $g_{1+n}\equiv I,\,n\geq1$, in initial data (\ref{BBGKYi}), then states
given in terms of the sequence of marginal density operators (\ref{RozvBBGKY}) can be described within
the framework of the sequence $F(t\mid F_{1}(t))=(1,F_1(t),F_2(t\mid F_{1}(t)),\ldots,F_s(t\mid F_{1}(t)),\ldots)$ of
explicitly defined functionals $F_s(t\mid F_{1}(t)),\,s\geq2$, of the solution $F_1(t)$ of the
generalized quantum kinetic equation.

We extend this statement to case of initial data (\ref{BBGKYi}) and construct the mean field
(self-consistent field) asymptotics of a solution of the Cauchy problem of the corresponding
generalized quantum kinetic equation.

\section{The generalized quantum kinetic equation in case of correlated initial state}
We reformulate the Cauchy problem (\ref{BBGKY})-(\ref{BBGKYi}) as the new Cauchy problem for
a one-particle density operator governed by the generalized quantum kinetic equation and the
sequence of explicitly defined marginal functionals of the state
$F_{s}\big(t,Y\mid F_{1}(t)\big),\,s\geq 2$, which are determined by the solution
$F_{1}(t)$ of such Cauchy problem. With this aim we introduce the following kinetic cluster
expansions of cumulants (\ref{cumulant}) of groups of operators (\ref{groupG})
\begin{eqnarray}\label{kcec}
   &&\hskip-5mm\mathfrak{A}_{1+n}(-t,\{Y\},X\setminus Y)g_{1+n}(\{Y\},X\setminus Y)
      \prod_{i=1}^{s+n}\mathfrak{A}_{1}(t,i)=\\
   &&\hskip-5mm=\sum_{n_1=0}^{n}\frac{n!}{(n-n_1)!}
      \mathfrak{G}_{1+n-n_1}(t,\{Y\},s+1,\ldots,s+n-n_1)\times\nonumber\\
   &&\hskip-5mm\times\sum_{\mathrm{D_{s+n}}:Z=\bigcup_i
      X_i}\,\,\sum_{i_1<i_2<\ldots<i_{|\mathrm{D_{s+n}}|}=1}^{s+n-n_1}
      \prod_{k=1}^{|\mathrm{D_{s+n}}|}\frac{1}{|X_k|!}\,
      \mathfrak{A}_{1+|X_{k}|}(-t,i_k,X_{k})\times\nonumber\\
   &&\hskip-5mm\times\prod_{k=1}^{|\mathrm{D_{s+n}}|}g_{1+|X_{k}|}(i_k,X_{k})\mathfrak{A}_{1}(t,i_k)
      \prod_{j\in Z}\mathfrak{A}_{1}(t,j),\nonumber
\end{eqnarray}
where $X\setminus Y\equiv(s+1,\ldots,s+n)$ and $\sum_{\mathrm{D_{s+n}}:Z=\bigcup_i X_i}$
is the sum over all possible dissections $\mathrm{D_{s+n}}$ of the linearly ordered set
$Z\equiv(s+n-n_1+1,\ldots,s+n)$ on no more than $s+n-n_1$ linearly ordered subsets.
We give a few examples of recurrence relations (\ref{kcec})
\begin{eqnarray*}
   &&\breve{\mathfrak{A}}_{1}(t,\{Y\})=\mathfrak{G}_{1}(t,\{Y\}),\\
   &&\breve{\mathfrak{A}}_{2}(t,\{Y\},s+1)=\mathfrak{G}_{2}(t,\{Y\},s+1)+
       \mathfrak{G}_{1}(t,\{Y\})\sum_{i_1=1}^s \breve{\mathfrak{A}}_{2}(t,i_1,s+1),\\
   &&\breve{\mathfrak{A}}_{3}(t,\{Y\},s+1,s+2)=\mathfrak{G}_{3}(t,\{Y\},s+1,s+2)+\\
   &&\hskip+8mm+2!\mathfrak{G}_{2}(t,\{Y\},s+1)\sum_{i_1=1}^{s+1}\breve{\mathfrak{A}}_{2}(t,i_1,s+2)+
       \mathfrak{G}_{1}(t,\{Y\})\big(\sum_{i_1=1}^s\breve{\mathfrak{A}}_{3}(t,i_1,s+1,s+2)+\\
   &&\hskip+8mm+2!\sum_{1=i_1<i_2}^s
       \breve{\mathfrak{A}}_{2}(t,i_1,s+1)\breve{\mathfrak{A}}_{2}(t,i_2,s+2)\big),
\end{eqnarray*}
where the operator $\widehat{\mathfrak{A}}_{1+n}(t)$ is the $(1+n)th$-order scattering cumulant
\begin{eqnarray}\label{sc}
   &&\hskip-5mm\breve{\mathfrak{A}}_{1+n}(t,\{Y\},X\backslash Y)\doteq
       \mathfrak{A}_{1+n}(-t,\{Y\},X\backslash Y)g_{1+n}(\{Y\},X\backslash Y)
       \prod_{i=1}^{s+n}\mathfrak{A}_{1}(t,i),
\end{eqnarray}

In terms of scattering cumulants (\ref{sc}) the solutions $\mathfrak{G}_{1+n}(t,\{Y\},X\backslash Y),\,n\geq0$,
of recurrence relations (\ref{kcec}) are represented by the following expansions
\begin{eqnarray}\label{skrrc}
   &&\hskip-8mm\mathfrak{G}_{1+n}(t,\{Y\},X\setminus Y)\doteq n!\,
       \sum_{k=0}^{n}\,(-1)^k\,\sum_{n_1=1}^{n}\ldots
       \sum_{n_k=1}^{n-n_1-\ldots-n_{k-1}}\frac{1}{(n-n_1-\ldots-n_k)!}\times\\
   &&\hskip-8mm\times\breve{\mathfrak{A}}_{1+n-n_1-\ldots-n_k}(t,\{Y\},s+1,\ldots,
       s+n-n_1-\ldots-n_k)\times\nonumber\\
   &&\hskip-8mm\times\prod_{j=1}^k\,\sum\limits_{\mbox{\scriptsize$\begin{array}{c}
       \mathrm{D}_{j}:Z_j=\bigcup_{l_j}X_{l_j},\\
       |\mathrm{D}_{j}|\leq s+n-n_1-\dots-n_j\end{array}$}}\frac{1}{|\mathrm{D}_{j}|!}
       \sum_{i_1\neq\ldots\neq i_{|\mathrm{D}_{j}|}=1}^{s+n-n_1-\ldots-n_j}\,\,
       \prod_{X_{l_j}\subset \mathrm{D}_{j}}\,\frac{1}{|X_{l_j}|!}\,\,
       \breve{\mathfrak{A}}_{1+|X_{l_j}|}(t,i_{l_j},X_{l_j}),\nonumber
\end{eqnarray}
where $\sum_{\mathrm{D}_{j}:Z_j=\bigcup_{l_j} X_{l_j}}$ is the sum over all possible dissections
of the linearly ordered set $Z_j\equiv(s+n-n_1-\ldots-n_j+1,\ldots,s+n-n_1-\ldots-n_{j-1})$ on
no more than $s+n-n_1-\ldots-n_j$ linearly ordered subsets. For example,
\begin{eqnarray*}
   &&\hskip-7mm\mathfrak{G}_{1}(t,\{Y\})=\breve{\mathfrak{A}}_{1}(t,\{Y\})=
       \mathfrak{A}_{1}(-t,\{Y\})g_{1}(\{Y\})\prod_{i=1}^{s}\mathfrak{A}_{1}(t,i),\\
   &&\hskip-7mm\mathfrak{G}_{2}(t,\{Y\},s+1)=\breve{\mathfrak{A}}_{2}(t,\{Y\},s+1)-
       \breve{\mathfrak{A}}_{1}(t,\{Y\})\sum_{i=1}^s\breve{\mathfrak{A}}_{2}(t,i,s+1),\\
   &&\hskip-7mm\mathfrak{G}_{3}(t,\{Y\},s+1,s+2)=\breve{\mathfrak{A}}_{3}(t,\{Y\},s+1,s+2)-\\
   &&-2!\breve{\mathfrak{A}}_{2}(t,\{Y\},s+1)\sum_{i_1=1}^{s+1}\breve{\mathfrak{A}}_{2}(t,i_1,s+2)-
       \breve{\mathfrak{A}}_{1}(t,\{Y\})\big(\sum_{i_1=1}^{s}\breve{\mathfrak{A}}_{3}(t,i_1,s+1,s+2)-\\
   &&-2!\sum_{i_1=1}^{s}\sum_{i_2=1}^{s+1}\widehat{\mathfrak{A}}_{2}(t,i_1,s+1)
       \breve{\mathfrak{A}}_{2}(t,i_2,s+2)+2!\sum_{1=i_1< i_2}^{s}\breve{\mathfrak{A}}_{2}(t,i_1,s+1)
       \breve{\mathfrak{A}}_{2}(t,i_2,s+2)\big).
\end{eqnarray*}

As a result of the application of expansion (\ref{kcec}) to solution (\ref{RozvBBGKY}) we establish
that the one-particle density operator $F_{1}(t)$ is governed by the following generalized quantum
kinetic equation
\begin{eqnarray}\label{gkec}
   &&\hskip-5mm\frac{d}{dt}F_{1}(t,1)=-\mathcal{N}(1)F_{1}(t,1)+\\
   &&\hskip-5mm+\epsilon\,\mathrm{Tr}_{2}(-\mathcal{N}_{\mathrm{int}}(1,2))
      \sum\limits_{n=0}^{\infty}\frac{1}{n!}\mathrm{Tr}_{3,\ldots,n+2}\,
      \mathfrak{G}_{1+n}(t,\{1,2\},3,\ldots,n+2)\prod _{i=1}^{n+2} F_{1}(t,i),\nonumber\\ \nonumber\\
   \label{vpgke}
   &&\hskip-5mmF_{1}(t)|_{t=0}=F_{1}^0.
\end{eqnarray}
where the operator $(-\mathcal{N}_{\mathrm{int}}(1,2))$ is defined by formula (\ref{comst}), and 
the $(1+n)th$-order generated evolution operators $\mathfrak{G}_{1+n}(t),\,n\geq0$, are determined 
by expansion (\ref{skrrc}) over scattering cumulants (\ref{sc}). In equation (\ref{gkec}) 
the collision integral series converges under the condition that \cite{GG}: 
$\|F_1(t)\|_{\mathfrak{L}^{1}(\mathcal{H})}<e^{-8}$.

The global in time solution of initial-value problem (\ref{gkec})-(\ref{vpgke}) is determined by
the following expansion
\begin{eqnarray}\label{ske}
    &&\hskip-7mmF_{1}(t,1)=\sum\limits_{n=0}^{\infty}\frac{1}{n!}\,\mathrm{Tr}_{2,\ldots,{1+n}}\,\,
        \mathfrak{A}_{1+n}(-t,1,2,\ldots,n+1)g_{1+n}(1,2,\ldots,n+1)\prod_{i=1}^{n+1}F_{1}^0(i),
\end{eqnarray}
where $\mathfrak{A}_{1+n}(-t)$ is the $(1+n)th$-order cumulant (\ref{cumulant}) of groups of operators
(\ref{groupG}). The series (\ref{ske}) converges under the condition that \cite{GG}:
$\|F_1^0\|_{\mathfrak{L}^{1}(\mathcal{H})}<e^{-10}(1+e^{-9})^{-1}$.

Correspondingly having applied expansion (\ref{kcec}) to solutions (\ref{RozvBBGKY}) in case of
$s\geq2$, in terms of solution expansion (\ref{ske}), they are represented as the following
marginal functionals of the state
\begin{eqnarray}\label{cf}
   &&\hskip-5mmF_{s}(t,Y\mid F_{1}(t))\doteq\sum_{n=0}^{\infty}\frac{1}{n!}\,
      \mathrm{Tr}_{s+1,\ldots,{s+n}}\,\mathfrak{G}_{1+n}(t,\{Y\},X\setminus Y)\prod_{i=1}^{s+n}F_{1}(t,i),
\end{eqnarray}
where the $(1+n)th$-order generated evolution operator $\mathfrak{G}_{1+n}(t)$ is determined by
expansion (\ref{skrrc}) over scattering cumulants (\ref{sc}). The series (\ref{cf}) converges
under the condition that: $\|F_{1}(t)\|_{\mathfrak{L}^{1}(\mathcal{H})}<e^{-(3s+2)}$. The constructed
marginal functionals of the state \eqref{cf} characterize the correlations of states of quantum
many-particle systems.

We note that the average values of the nonadditive-type marginal observables are determined in
terms of marginal functionals of the state (\ref{cf}) and in case of the additive-type marginal
observables they are determined by solution (\ref{ske}) of the generalized quantum kinetic
equation (\ref{gkec}).

Thus, in case of initial data (\ref{BBGKYi}) solution (\ref{RozvBBGKY}) of the Cauchy problem
(\ref{BBGKY})-(\ref{BBGKYi}) of the quantum BBGKY hierarchy and a solution of the Cauchy problem
of the generalized kinetic equation (\ref{gkec})-(\ref{vpgke}) together with marginal functionals
of the state (\ref{cf}) give two equivalent approaches to the description of the evolution of
quantum many-particle systems. The coefficients of generalized quantum kinetic equation (\ref{gkec})
and generated evolution operators (\ref{skrrc}) of marginal functionals (\ref{cf}) are determined
by the operators of initial correlations.

\section{The mean field limit theorems}
In case of initial state involving correlations for an asymptotic perturbation of generated
evolution operator (\ref{skrrc}) in the mean field limit the following equality is valid
\begin{eqnarray*}
  &&\lim\limits_{\epsilon\rightarrow 0}\big\|\mathfrak{G}_{1+n}(t,\{Y\},X\setminus Y)
     f_{s+n}\big\|_{\mathfrak{L}^{1}(\mathcal{H}_{s+n})}=0, \quad n\geq1,
\end{eqnarray*}
and in case of the first-order generated evolution operator (\ref{skrrc}) we have respectively
\begin{eqnarray*}
  &&\lim\limits_{\epsilon\rightarrow 0}\big\|\big(\mathfrak{G}_{1}(t,\{Y\})-
     \prod_{i_1=1}^{s}\mathcal{G}_{1}(-t,i_1)g_{1}(\{Y\})\prod_{i_2=1}^{s}\mathcal{G}_{1}(t,i_2)\big)
     f_{s}\big\|_{\mathfrak{L}^{1}(\mathcal{H}_{s})}=0.
\end{eqnarray*}

In view that under the condition that:
$t<t_{0}\equiv\big(2\,\|\Phi\|_{\mathfrak{L}(\mathcal{H}_{2})}\|\epsilon\,F_{1}^0\|_{\mathfrak{L}^{1}(\mathcal{H})}\big)^{-1}$,
the series for $\epsilon\,F_1(t)$ converges, then for $t<t_0$ the remainders of solution series (\ref{ske}) can be 
made arbitrary small for sufficient large $n=n_0$ independently of $\epsilon$. Then, using stated above asymptotic
perturbation formulas for each integer $n$ every term of this series converges term by term to the limit operator
$f_{1}(t)$ which is represented in the form of the following expansion
\begin{eqnarray}\label{viter}
    &&\hskip-8mmf_{1}(t,1)=\sum\limits_{n=0}^{\infty}\int\limits_0^tdt_{1}\ldots\int\limits_0^{t_{n-1}}dt_{n}
        \mathrm{Tr}_{\mathrm{2,\ldots,1+n}}\mathcal{G}_{1}(-t+t_{1},1)
        \big(-\mathcal{N}_{\mathrm{int}}(1,2)\big)\times\\
    &&\hskip-8mm\times\prod\limits_{j_1=1}^{2}\mathcal{G}_{1}(-t_{1}+t_{2},j_1)\ldots
        \prod\limits_{j_{n-1}=1}^{n}\mathcal{G}_{1}(-t_{n-1}+t_{n},j_{n-1})\times\nonumber\\
    &&\hskip-8mm\times \sum\limits_{i_{n}=1}^{n}\big(-\mathcal{N}_{\mathrm{int}}(i_{n},1+n)\big)
        \prod\limits_{j_n=1}^{1+n}\mathcal{G}_{1}(-t_{n},j_n)g_{1+n}(1,\ldots,n+1)
        \prod\limits_{i=1}^{1+n}f_{1}^0(i)\nonumber.
\end{eqnarray}
For bounded interaction potentials series (\ref{viter}) is norm convergent on the space
$\mathfrak{L}^{1}(\mathcal{H})$ under the condition that: $t<t_{0}\equiv\big(2\,\|\Phi\|_{\mathfrak{L}(\mathcal{H}_{2})}
\|f_1^0\|_{\mathfrak{L}^{1}(\mathcal{H})}\big)^{-1}$.

Thus, if there exists the limit $f_{1}^0\in\mathfrak{L}^{1}(\mathcal{H})$ of initial data (\ref{vpgke}), namely
\begin{eqnarray*}
  &&\lim\limits_{\epsilon\rightarrow 0}\big\|\epsilon\,F_{1}^0-f_{1}^0\big\|_{\mathfrak{L}^{1}(\mathcal{H})}=0,
\end{eqnarray*}
then for finite time interval $t\in(-t_{0},t_{0}),$ where
$t_{0}\equiv\big(2\,\|\Phi\|_{\mathfrak{L}(\mathcal{H}_{2})}\|f_1^0\|_{\mathfrak{L}^{1}(\mathcal{H})}\big)^{-1},$
there exists the mean field limit of solution expansion (\ref{ske}) of the generalized quantum kinetic equation (\ref{gkec}):
\begin{eqnarray}\label{1lim}
    &&\hskip-5mm\lim\limits_{\epsilon\rightarrow 0}\big\|\epsilon\,F_{1}(t)-
       f_{1}(t)\big\|_{\mathfrak{L}^{1}(\mathcal{H})}=0,
\end{eqnarray}
where the operator $f_{1}(t)$ is represented by series (\ref{viter}) and it is a solution
of the Cauchy problem of the modified quantum Vlasov kinetic equation
\begin{eqnarray}\label{mVe}
  &&\hskip-5mm\frac{d}{dt}f_{1}(t,1)=-\mathcal{N}(1)f_{1}(t,1)+\\
  &&\hskip-5mm+\mathrm{Tr}_{2}(-\mathcal{N}_{\mathrm{int}}(1,2))
     \prod_{i_1=1}^{2}\mathcal{G}_{1}(-t,i_1)g_{1}(\{1,2\})
     \prod_{i_2=1}^{2}\mathcal{G}_{1}(t,i_2)f_{1}(t,1)f_{1}(t,2),\nonumber\\ \nonumber\\
   \label{Vlasov2}
  &&\hskip-5mmf_{1}(t)|_{t=0}=f_{1}^0.
\end{eqnarray}

Since a solution of initial-value problem (\ref{gkec})-(\ref{vpgke}) of the generalized kinetic
equation converges to a solution of initial-value problem (\ref{mVe})-(\ref{Vlasov2}) of the
modified quantum Vlasov kinetic equation as (\ref{1lim}), for marginal functionals of the state 
(\ref{cf}) we correspondingly establish
\begin{eqnarray}\label{lf}
  &&\hskip-7mm\lim\limits_{\epsilon\rightarrow 0}\big\|\epsilon^{s}F_{s}(t,Y\mid F_{1}(t))-
     \prod _{i_1=1}^{s}\mathcal{G}_{1}(-t,i_1)g_{1}(\{Y\})\prod_{i_2=1}^{s}\mathcal{G}_{1}(t,i_2)
     \prod\limits_{j=1}^{s}f_{1}(t,j)\big\|_{\mathfrak{L}^{1}(\mathcal{H}_{s})}=0.
\end{eqnarray}
This equality means the propagation of initial correlations in time in the mean field limit.

\section{On mean field quantum kinetic equations}
Let us consider the pure states, i.e. the operator $f_{1}(t)=|\psi_{t}\rangle\langle\psi_{t}|$
is a one-dimensional projector onto a unit vector $|\psi_{t}\rangle\in\mathcal{H}$ and its kernel 
has the following form: $f_{1}(t,q,q')=\psi(t,q)\psi(t,q')$. 
Then, in case of initial data, satisfying a chaos property, it holds
\begin{eqnarray*}
   &&\lim\limits_{\epsilon\rightarrow 0}\big\|\epsilon\,F_{1}^0-
     |\psi_{0}\rangle\langle\psi_{0}|\big\|_{\mathfrak{L}^{1}(\mathcal{H})}=0,
\end{eqnarray*}
and in this case statement (\ref{lf}) reads
\begin{eqnarray*}
   &&\lim\limits_{\epsilon\rightarrow 0}\big\|\,\epsilon^{s} F_{s}\big(t \mid F_{1}(t)\big)-
     \prod _{i_1=1}^{s}\mathcal{G}_{1}(-t,i_1)g_{1}(\{Y\})\prod_{i_2=1}^{s}\mathcal{G}_{1}(t,i_2)
     |\psi_{t}\rangle\langle\psi_{t}|^{\otimes s}\,\big\|_{\mathfrak{L}^{1}(\mathcal{H}_{s})}=0,
\end{eqnarray*}
where the function $|\psi_{t}\rangle\in\mathcal{H}$ is a solution of the Cauchy problem of the
nonlinear Hartree equation for initial data $|\psi_{0}\rangle$. In particular, we remark that 
in case of a system of particles, interacting by the potential which kernel is the Dirac
measure, it reduces to the cubic nonlinear Schr\"{o}dinger equation
\begin{eqnarray*}
   &&i\frac{\partial}{\partial t}\psi(t,q)=-\frac{1}{2}\Delta_{q}\psi(t,q)+|\psi(t,q)|^{2}\psi(t,q).
\end{eqnarray*}

In case of correlated initial state (\ref{BBGKYi}) the sufficient equation for the description of 
the pure state evolution governed by modified quantum Vlasov kinetic equation (\ref{mVe}) is the 
Gross-Pitaevskii-type kinetic equation
\begin{eqnarray}\label{gpeq}
  &&\hskip-8mmi\frac{\partial}{\partial t}\psi(t,q)=-\frac{1}{2}\Delta_{q}\psi(t,q)+
     \int d q'd q''\mathfrak{b}(t,q,q;q',q'')\psi(t,q'')\psi^{\ast}(t,q)\psi(t,q'),
\end{eqnarray}
where the coupling ratio function $\mathfrak{b}(t,q,q;q',q'')$ is the kernel of the scattering length 
operator: $\prod_{i_1=1}^{2}\mathcal{G}_{1}(-t,i_1)b_{1}(\{1,2\})\prod_{i_2=1}^{2}\mathcal{G}_{1}(t,i_2)$.

Observing that in the macroscopic scale of the variation of variables, groups of operators
(\ref{groupG}) of finitely many particles depend on microscopic time variable $\varepsilon^{-1}t$,
where $\varepsilon\geq0$ is a scale parameter, the dimensionless marginal functionals of the state
are represented in the form: $F_{s}\big(\varepsilon^{-1}t\mid F_{1}(t)\big)$. As a result of the
limit processing $\varepsilon\rightarrow0$ we establish the Markovian kinetic evolution (\ref{gpeq})
with the corresponding coefficient $\mathfrak{b}(\varepsilon^{-1}t)$.

\section{Conclusion}
We generalized the concept of quantum kinetic equations for the kinetic evolution
involving correlations of particle states at initial time \cite{CGP97}, for instance,
correlations characterizing the condensed states \cite{BQ}. The mean field scaling
asymptotics of a solution of the generalized quantum kinetic equation of many-particle
systems in condensed states was analyzed. These results can be extended to quantum
systems of bosons or fermions \cite{GP}.

It should be emphasized that the kinetic evolution is an inherent property of infinite-particle
systems. In spite of the fact that in terms of a one-particle marginal density operator from the
space of trace-class operators can be described a system with the finite average number of particles,
the generalized quantum kinetic equation has been derived on the basis of the formalism of
nonequilibrium grand canonical ensemble since its framework is adopted to the description of
infinite-particle systems in suitable Banach spaces as well.

We note that one more approach to the construction of the kinetic equations in the mean field limit
in case of the presence of correlations at initial time is based on the description of the kinetic
evolution in terms of the evolution of marginal observables \cite{BG,G11}.

We remark that developed approach is also related to the problem of a rigorous derivation of
the non-Markovian kinetic-type equations from underlaying many-particle dynamics which make
possible to describe the memory effects of particle.

\bigskip

\addcontentsline{toc}{section}{References}
\renewcommand{\refname}{References}

\end{document}